\documentclass[twocolumn,tighten]{aastex631}
\hypersetup{linkcolor=blue,citecolor=blue,filecolor=blue,urlcolor=blue}
\usepackage{amsmath}
\usepackage{natbib}
\usepackage{xspace}
\usepackage{hyperref}

\newcommand{\feh}{{\mathrm{[Fe/H] }}}

\newcommand{\degree}{$^{\circ}$}
\newcommand{\loggf}{\mbox{$\log gf$}}
\newcommand{\kmsec}{\mbox{km~s$^{\rm -1}$}}
\newcommand{\logg}{\mbox{log~{\it g}}\xspace}
\newcommand{\msun}{\mbox{$M_{\odot}$}}
\newcommand{\teff}{\mbox{$T_{\rm eff}$}\xspace}
\newcommand{\vt}{\mbox{$v_{\rm t}$}}

\shorttitle{New parameters and abundances for 311 stars}
\shortauthors{Mittal \& Roederer}

\DeclareUnicodeCharacter{2212}{-}

\begin{document}

\title{%
New Stellar Parameters, Metallicities, and Elemental Abundance Ratios for 311 Metal-Poor Stars}

\author[0009-0000-9943-3524]{Sanil Mittal}
\affiliation{%
Department of Astronomy, University of Michigan,
1085 S.\ University Ave., Ann Arbor, MI 48109, USA}
\email{Email:\ sanilm@umich.edu}

\author[0000-0001-5107-8930]{Ian U.\ Roederer}
\affiliation{%
Department of Physics, North Carolina State University,
Raleigh, NC 27695, USA}
\affiliation{%
Department of Astronomy, University of Michigan,
1085 S.\ University Ave., Ann Arbor, MI 48109, USA}

\affiliation{%
Joint Institute for Nuclear Astrophysics -- Center for the
Evolution of the Elements (JINA-CEE), USA}

\begin{abstract}
We present equivalent widths, improved model atmosphere parameters, and revised abundances for 14 species of 11 elements derived from high resolution optical spectroscopy of 311 metal-poor stars. All of these stars had their parameters previously published by \citeauthor{roederer14c} We use color-\teff relationships calibrated for Gaia and 2MASS photometry to calculate improved effective temperatures (\teff). We calculate log of surface gravity (\logg) values using measurements derived from Gaia parallaxes and other fundamental stellar properties. We perform a standard LTE abundance analysis using MARCS model atmospheres and the MOOG line analysis software to rederive microturbulence velocity parameters, metallicities, and abundances based on O~\textsc{i}, Na~\textsc{i}, Mg~\textsc{i}, Si~\textsc{i}, K~\textsc{i}, Ca~\textsc{i}, Ti~\textsc{i}, Ti~\textsc{ii}, Cr~\textsc{i}, Cr~\textsc{ii}, Fe~\textsc{i}, Fe~\textsc{ii}, Ni~\textsc{i}, and Zn~\textsc{i} lines using the previously measured equivalent widths. On average, the new \teff values are 310 K warmer, the new \logg values are higher by 0.64 dex, and the new $\feh$ values are higher by 0.26 dex. We apply NLTE corrections to the abundances derived from O~\textsc{i}, Na~\textsc{i}, Mg~\textsc{i}, Si~\textsc{i}, K~\textsc{i}, Fe~\textsc{i}, and Fe~\textsc{ii} lines. Our sample contains 6 stars with $\feh < −3.5$, 28 stars with $\feh < −3.0$, and 113 stars with $\feh < −2.5$. Our revised abundances for these 311 stars are now in better agreement with those derived by previous studies of smaller samples of metal-poor stars in the Milky Way. 

\end{abstract}


\section{Introduction}
\label{intro}

Stars synthesize heavy elements through various channels during their lifetime and eject them through stellar winds or supernovae explosions into the nearby interstellar medium (ISM). The next generation of stars is formed in these gas clouds, retaining the heavier elements formed in the previous stellar generations. At the ends of their lifetimes, these stars also return their synthesized elements to the ISM.~ This cycle began with the first stars, which formed only a few hundred million years after the Big Bang, and has continued until today (e.g., \citealp{beers_christlieb05}; \citealp{nomoto_etal13}; \citealp{frebel15araa}). Today’s stars act as fossils, retaining information about the properties of the stars that came before them. Therefore, studying the elemental abundances of the present stellar populations in galaxies provides insight into their star formation history and evolution similar to Galactic archaeology (\citealt{kobayashi2020}).

Chemical abundances of stars can be derived from analyzing the absorption lines in the spectra. To perform this analysis, a model of the stellar atmosphere is needed, which requires the effective surface temperature (\teff) and the star’s surface gravity (\logg) values as inputs. Many studies have done this for hundreds of metal-poor stars identified in our Galaxy (e.g., \citealp{cayrel04}; \citealp{roederer14c}; \citealp{jacobson15smss}).

\cite{roederer14c} derived detailed chemical abundances for 53 species of 48 elements in each of 313 metal-poor stars in the Galaxy. That study derived \teff and \logg values directly from the spectra of each star, because the sample itself was drawn from several input catalogs with inhomogeneous auxiliary data, such as broadband photometry or parallax measurements.  In contrast, most other studies of metal-poor stars focused on smaller samples drawn primarily from one catalog each (e.g., \citealp{cayrel04,lai08,jacobson15smss}). Those studies generally relied on \teff and \logg values calculated from photometric color-\teff relations, which typically yield warmer \teff values than spectroscopic methods (e.g., \citealp{johnson02abund,jofre2019}). The systematically different \teff scale inevitably leads to systematically different \logg and microturbulent velocity parameter (\vt) values, and the derived chemical abundances are also affected.  Fortunately, ratios among the abundance of one element to another are less sensitive to these differences, but \citeauthor{roederer14c} stated in their section~9.4.1 that their [X/Fe] abundance ratios were usually overabundant in comparison to other surveys.

With the recent availability of all-sky Gaia photometry (\citealt{gaia2016, gaiadr3}), the process of inferring a \teff using photometry has been made easier for virtually all relatively bright ($V < 16$) stars in the sky. \cite{mucciarelli2021} presented a metallicity-dependent color-\teff transformation based on photometry from the early third data release (EDR3) of the Gaia mission and the near-infrared Two Micron All-Sky Survey (2MASS) photometric survey (\citealt{2mass2006}). They build on the commonly used infrared flux method to calculate a \teff by calibrating the flux measured in different Gaia passbands to an effective temperature by assuming a metallicity.

In this study, we use Gaia and 2MASS photometry to calculate new \teff values for the stars in the \cite{roederer14c} sample. We also calculate \logg from the fundamental physical relations. By using these new stellar parameter estimates, along with equivalent widths (EWs) measured previously from high-resolution spectroscopy, we rederive new abundances for O, Na, Mg, Si, K, Ca, Ti, Cr, Fe, Ni, and Zn for the stars initially published by \citeauthor{roederer14c} We also include non-local thermodynamic equilibrium (NLTE) corrections for the abundances of O, Na, Mg, Si, K, and Fe. Detailed abundances for elements requiring spectrum synthesis, such as V, Y, Zr, Eu, etc., are beyond the scope of this paper and will be presented separately. 

We introduce the adopted input data in Section~\ref{obs}. We describe our stellar parameter and metallicity calculations in Section~\ref{params}, and we describe our derivation of abundances for other elements in Section~\ref{abundances}. We share our results in Section~\ref{results}. We compare our abundances with other stellar samples in Section~\ref{discussion}. Finally, we summarize our conclusions in Section~\ref{conclusions}.

\section{Observations and Data}
\label{obs}

We started with the sample published by \citet{roederer14c}. We used the CDS XMatch (\citealt{cds2020}) service to cross-reference these stars with the Gaia mission's third data release (DR3; \citealt{gaia2016, gaiadr3}), near-infrared Two Micron All-Sky Survey (2MASS) photometric survey (\citealt{2mass2006}), the Gaia Early Data Release 3 (EDR3) Corrected Distances catalog \citet{bailerjones2021}, and the reddening estimates from \citet{schlafly11} available through the NASA/IPAC Infrared Science Archive. All stars in this sample are relatively bright compared to their neighboring stars, so the matches are unambiguous, and we confirmed each one manually. A few stars returned multiple potential matches, and we resolved these situations by manual inspection.

To derive abundances, we need estimates of stellar parameters, such as \teff, \logg, \vt, and an estimate of the star's metallicity. The derivation of \teff (Section~\ref{temperature}) requires an estimate of reddening to correct the observed colors, and the reddening requires a distance estimate for each star. Calculating \logg also requires a distance and an estimate of the stellar mass (Section~\ref{gravity}).

For most of the stars, we obtained a distance estimate using the Gaia EDR3 Corrected Distances catalog \citep{bailerjones2021}. Instead of simply inverting a parallax to obtain a distance, \citeauthor{bailerjones2021}\ derived distances using a probabilistic approach based on a three-dimensional model of the Milky Way. Their model includes interstellar extinction and Gaia's variable magnitude limit. We adopt the photogeometric distances they provide.

We also retrieved reddening values along the line of sight given by the dust maps presented by \citet{schlafly11}. These dust maps assume the source is located behind the reddening layer, so we correct them to account for only the dust in the foreground of each star. If the distance to the star is $< 70$~pc, we adopt a reddening, $E(B-V)$, of 0.00 \citep{casagrande10}. Otherwise, we correct for the reddening to the star at a distance $d$ using the following equation \citep{bonifacio00phot}:
\begin{equation}
    \begin{split}
        E(B-V) = E(B-V)_{\rm inf} (1 - e^{-\lvert d \sin(b)\rvert / h})
    \end{split}
\end{equation}
where $b$ is the Galactic latitude and $E(B-V)_{\rm inf}$ is the reddening at infinity. The scale height $h$ for the reddening layer is assumed to be 125~pc (\citeauthor{bonifacio00phot}).

Finally, we remove two stars from the sample, \mbox{CS~30339-046} and \mbox{CS~30312-062}.
\cite{clementini2023} lists \mbox{CS~30339-046} and \mbox{CS~30312-062} as RR~Lyr variables, so we discard them from further consideration in our study.

\section{Stellar Parameters}
\label{params}

Here we discuss our calculations performed to obtain the values for the individual stellar parameters:\ 
\teff, \logg, \vt, and 
model metallicity ([M/H]). We use Model Atmospheres with a Radiative and Convective Scheme (MARCS; \citealt{gustafsson08}), MOOG \citep{sneden73}, and various atomic and molecular spectral line data to generate synthetic stellar spectra. MARCS is a grid of one-dimensional (1-D), hydrostatic, plane-parallel and spherical model atmospheres calculated under the assumption of local thermodynamic equilibrium (LTE).~ MOOG performs a variety of LTE line analysis and spectrum synthesis tasks, including the determination of chemical composition of stars.

\subsection{Effective Temperature}
\label{temperature}

We calculate \teff using the 
metallicity-dependent color-\teff\ relations presented in
\citet{mucciarelli2021}. That study presented 
six color-\teff\ relations constructed from four photometric bands:\
$BP-RP$, $G-RP$, $BP-G$, $G-K$, $BP-K$, and $RP-K$. We adopt the $G$, $BP$, and $RP$ photometry from Gaia, and the $K$ photometry from 2MASS.
We estimate the reddening, $E(B-V)$, using the dust maps presented by \citet{schlafly11}. We deredden the photometry using the extinction coefficient presented by \citet{mccall04} for $K$, and we iteratively calculate the color-dependent coefficients for $G$, $BP$, and $RP$ using the formulae presented by \citet{gaia2018c}. The reddening is small in all cases, with a median $E(B-V)$ of 0.034. 

We adopt the \citet{roederer14c} \logg and [M/H] values as our initial estimates for these values. We use an analogous method to that described in \citet{roederer18c} to calculate \teff for each star. We draw $10^4$ samples from each input parameter (metallicity, $G$, $BP$, $RP$, $K$, parallax, distance, and reddening) and calculate the \teff value predicted by each set of input values. Each calculation uses the same set of input draws consistently throughout. We adopt the median of the \teff distribution for each color, and we combine the six \teff predictions using an average that is weighted by the uncertainty in the six bands to calculate a final \teff value.
Table~\ref{tefftab} lists the \teff\ values calculated from each of the six colors.

\begin{deluxetable*}{ccccccccccccc}
\tablecaption{Effective temperatures calculated from different colors
\label{tefftab}}
\tablehead{
\colhead{Star} &
\colhead{($BP-RP$)} &
\colhead{$\sigma$} &
\colhead{($BP-G$)} &
\colhead{$\sigma$} &
\colhead{($G-RP$)} &
\colhead{$\sigma$} &
\colhead{($BP-K$)} &
\colhead{$\sigma$} &
\colhead{($G-K$)} &
\colhead{$\sigma$} &
\colhead{($RP-K$)} &
\colhead{$\sigma$} 
\\
\colhead{} &
\colhead{(K)} &
\colhead{(K)} &
\colhead{(K)} &
\colhead{(K)} &
\colhead{(K)} &
\colhead{(K)} &
\colhead{(K)} &
\colhead{(K)} &
\colhead{(K)} &
\colhead{(K)} &
\colhead{(K)} &
\colhead{(K)} 
}
\startdata
BD+10\degree2495  & 5110 & 89 & 5115 & 90 & 5107 & 80 & 5076 & 60 & 5041 & 78  & 5061 & 61\\ 
BD+19\degree1185A & 5546 & 68 & 5579 & 70 & 5524 & 73 & 5506 & 57 & 5477 & 77 & 5492 & 64\\  
BD+24\degree1676  & 6538 & 76 & 6516 & 79 & 6540 & 83 & 6506 & 68 & 6478 & 93 & 6493 & 75\\  
BD+26\degree3578  & 6685 & 70 & 6681 & 76 & 6668 & 77 & 6680 & 74 & 6671 & 110 & 6666 & 84\\
BD+29\degree2356  & 4927 & 88 & 4931 & 89 & 4924 & 78 & 4913 & 58 & 4897 & 75 & 4903 & 58\\
\enddata
\tablecomments{%
The complete version of Table~\ref{tefftab} is available in the online edition
of the journal.
A short version is included here to demonstrate its form and content.}

\end{deluxetable*}

For two of the stars (\mbox{G190-010} and \mbox{HD~219617}), the Gaia $G$ bandpass photometry results in values that disagree with the other bands at the $\approx$100 K level. For these stars, we only consider the three colors that do not make use of the $G$ bandpass. 

We calculate the uncertainty in \teff~from each color by adding the standard deviation ($\sigma$) of all $10^4$ values in quadrature with the $\sigma_{\teff}$ value given for each color in table~1 of \cite{mucciarelli2021}. This latter uncertainty accounts for the $1\sigma$ dispersion around the polynomial fit derived by \citeauthor{mucciarelli2021} We combine the uncertainties in \teff of the individual colors to calculate the uncertainties and the \teff values presented in Table~\ref{phottab}. To calculate the uncertainty in the adopted \teff value, we first calculate the $\sigma_f$ value, which is defined as the inverse square root of the sum of the square of the uncertainties in the \teff of each color used for a star. We then calculate the $\sigma_n$ value, defined as the standard deviation of the individual color's \teff values, where $n$ is the number of colors used. We then combine these two uncertainties as follows:

\begin{equation}
    \sigma_{\rm final} = \sqrt{\sigma_f^2 + \frac{\sigma_n^2}{n}}
\end{equation}
where $\sigma_{\rm final}$ is the uncertainty in \teff reported for each star.
\subsection{Surface Gravity}
\label{gravity}

We calculate \logg by using the same method as described in section~3.2 of \citet{roederer18c}, which used the fundamental relation:
\begin{equation}
\label{logg_eqn}
\begin{split}
    \logg & = 4 \log \teff + \log (M/\msun) - 10.61 + 0.4(BC_G \\ &+ G + 5 \log \varpi + 5 - R_GE(B - V) - M_{bol,\odot})
\end{split}
\end{equation}
Here, $\varpi$ represents the Gaia parallax, but for most stars we adopted the EDR3 distance given by \citet{bailerjones2021}, where we replaced $\varpi$ with $1/d$.

For five stars (\mbox{CS~22871-104}, \mbox{CS~22940-070}, \mbox{CS~22963-004}, \mbox{CS~29509-027}, and \mbox{HD~219617}), we use the Gaia parallax, rather than the \citet{bailerjones2021} distances, because the calculated \logg value lies closer to the expected main sequence or horizontal branch track in Figure~\ref{tefflogg}, and these stars have very small uncertainties in parallax. For example, \mbox{CS~22963-004} has \teff = 5603 $\pm$ 29~K, and its \logg = 3.03 $\pm$ 0.22 when calculated from the \citeauthor{bailerjones2021}\ distance and 2.47 $\pm$ 0.24 when calculated directly from the Gaia parallax. Using the \logg\ value calculated from its distance places the star in a region between the horizontal branch and the subgiants, whereas the \logg value calculated from the parallax places it squarely on the horizontal branch.

In Equation~\ref{logg_eqn}, $M_{bol,\odot}$ is the solar bolometric magnitude, 4.75, and the constant 10.61 is calculated from the solar constants log {\teff,}$_\odot$ = 3.7617 and $\logg_\odot$ = 4.438. The mass, $M$, of each star, was taken to be 0.8 \msun\ with an uncertainty of 0.1 \msun. $BC_G$ is the bolometric correction, given by \citet{casagrande2018a, casagrande2018b}. The constant $k_G$ is calculated using equation~1 and table~1 of \citet{gaia2018c}, and $R_G = 3.1k_G$. 

We draw $10^4$ samples for each of the input parameters (\teff, $BC_G$, $G$, $E(B - V)$, $M$, and $d$ or $\varpi$) from a normal distribution with the value as the mean and the uncertainty in each parameter as the standard deviation of the distribution. From the $10^4$ values obtained for the \logg value, we adopt the median of these calculations as the $\logg$ value, and we adopt the standard deviation as its uncertainty. 
 
\subsection{Microturbulent Velocity and Model Metallicity}
\label{microturbulent}

We adopt the standard approach to estimate \vt\ and [M/H], with the use of abundances derived from Fe~\textsc{i} lines (for \vt) and Fe~\textsc{ii} lines (for [M/H]). We adopted EWs measured previously by I.U.R., but not previously published as part of the \citet{roederer14c} study. We adopt the Fe~\textsc{ii} abundance rounded to the nearest 0.1 as [M/H].
These Fe~\textsc{i} and \textsc{ii} lines are listed in 
Table~\ref{abundtab}.

\begin{deluxetable*}{cccccccccccccc}
\tablecaption{Derived line-by-line Abundances
\label{abundtab}}
\tablewidth{0pt}
\tabletypesize{\small}
\tablehead{
\colhead{Star} &
\colhead{Species} &
\colhead{$\lambda$} &
\colhead{EW} &
\colhead{$\log\epsilon$(X)} &
\colhead{Corr}\\
\colhead{} &
\colhead{} &
\colhead{(\AA)} &
\colhead{} &
\colhead{LTE} &
\colhead{NLTE}
}
\startdata
\mbox{BD+10\degree2495} & Fe~\textsc{i} & 3805.34 & 83.7 & 5.33 & ... \\
\mbox{BD+10\degree2495} & Fe~\textsc{i} & 3949.95 & 76.3 & 5.41 & ... \\
\mbox{BD+10\degree2495} & Fe~\textsc{i} & 4147.67 & 79.7 & 5.45 & ... \\
\mbox{BD+10\degree2495} & Fe~\textsc{i} & 4154.50 & 74.9 & 5.49 & ... \\
\mbox{BD+10\degree2495} & Fe~\textsc{i} & 4157.78 & 57.2 & 5.41 & ... \\
\enddata      
\tablecomments{%
The complete version of Table~\ref{abundtab} is available in the online edition of the journal. A short version is included here to demonstrate its form and content.
}

\end{deluxetable*}

We use the stellar parameter values provided in \citet{roederer14c} as the initial estimates for \vt\ and [M/H], and we interpolate MARCS model atmospheres \citep{gustafsson08} using a code kindly provided by A. McWilliam (private communication 2009). 
We derive Fe abundances 
using the ``abfind'' driver of a recent\footnote{\url{https://github.com/jsobeck/MOOG_SCAT}} version of the line analysis software MOOG
(\citealt{sneden73}), which
assumes LTE.~
This version of MOOG treats
Rayleigh scattering, which affects 
the continuous opacity at shorter wavelengths,
as isotropic, coherent scattering
\citep{sobeck11}.
We adopt damping constants for collisional broadening
with neutral hydrogen from \citet{barklem00h}
and \citet{barklem05feii}, when available,
otherwise
we adopt the standard \citet{unsold55} recipe. We determine the \vt\ value for each model that minimizes the correlation between
the abundance derived from Fe~\textsc{i} lines
and line strength.
We adopt the [M/H] value for each model that matches the iron 
(Fe, $Z =$~26) abundance derived from the Fe~\textsc{ii} lines rounded to the nearest 0.1.
With this updated metallicity, we repeat the steps in Section~\ref{temperature} and Section~\ref{gravity} to iteratively determine \teff, \logg, \vt, and [M/H] until we reach convergence. The criterion for convergence was adopted as a change of $<10$~K in \teff. The final model atmosphere parameters are reported in Table~\ref{phottab}.

\begin{deluxetable*}{lcccccccccccccc}
\tablecaption{Final \teff\ and \logg\ Values, and Other Final Model Atmosphere Parameters
\label{phottab}}
\tablewidth{0pt}
\tabletypesize{\scriptsize}
\tablehead{
\colhead{Star} &
\colhead{Class\tablenotemark{a}} &
\colhead{\teff} &
\colhead{Unc.} &
\colhead{\logg} &
\colhead{Unc.} &
\colhead{$E(B-V)$} &
\colhead{Unc.} &
\colhead{$BC_{G}$} &
\colhead{Unc.} &
\colhead{[M/H]} &
\colhead{Unc.} &
\colhead{\vt} & 
\colhead{Unc.}\\
\colhead{} &
\colhead{} &
\colhead{} &
\colhead{\teff} &
\colhead{} &
\colhead{\logg} &
\colhead{} &
\colhead{$E(B-V)$} &
\colhead{} &
\colhead{$BC_{G}$} &
\colhead{} &
\colhead{[M/H]} &
\colhead{} &
\colhead{\vt}\\
\colhead{} &
\colhead{} &
\colhead{(K)} &
\colhead{(K)} &
\colhead{[cgs]} &
\colhead{[cgs]} &
\colhead{(mag)} &
\colhead{(mag)} &
\colhead{(mag)} &
\colhead{(mag)} &
\colhead{} &
\colhead{(dex)} &
\colhead{(\kmsec)} & 
\colhead{(\kmsec)}
}
\startdata
BD+10\degree2495 & RG & 5079 & 29 & 2.47 & 0.16 & 0.02 & 0.01 & $-$0.09 & 0.01 & $-$1.9 & 0.3 & 1.55 & 0.06 \\
BD+19\degree1185A & MS & 5520 & 36 & 4.35 & 0.16 & 0.00 & 0.01 & 0.02 & 0.01 & $-$1.2 & 0.3 & 1.15 & 0.06\\
BD+24\degree1676 & MS & 6515 & 36 & 4.05 & 0.15 & 0.02 & 0.01 & $-$0.09 & 0.01 & $-$2.3 & 0.3 & 1.65 & 0.06 \\
BD+26\degree3578 & MS & 6677 & 36 & 4.13 & 0.15 & 0.05 & 0.01 & $-$0.14 & 0.01 & $-$2.1 & 0.3 & 1.70 & 0.06 \\
BD+29\degree2356 & RG & 4913 & 29 & 1.71 & 0.19 & 0.01 & 0.01 & $-$0.09 & 0.01 & $-$1.5 & 0.3 & 1.68 & 0.06
\enddata      
\tablecomments{%
The complete version of Table~\ref{phottab} is available in the online edition of the journal. A short version is included here to demonstrate its form and content.}
\tablenotetext{a}{The classification scheme is defined as follows: BS, blue straggler-like stars warmer than the main sequence turnoff; HB, stars on the horizontal branch; MS, stars on the main sequence; RG, stars on the red giant branch; SG, stars on the subgiant branch.}
\end{deluxetable*}

\section{Abundance Analysis}
\label{abundances}

We adopt the standard nomenclature for elemental abundances and ratios. The abundance of element X is defined as the number of X atoms per 10$^{12}$ H atoms, $\log\epsilon$(X)~$\equiv \log_{10}(N_{\rm X}/N_{\rm H})+$12.0. The ratio of the abundances of elements X and Y relative to the Solar ratio is defined as [X/Y] $\equiv \log_{10} (N_{\rm X}/N_{\rm Y}) - \log_{10} (N_{\rm X}/N_{\rm Y})_{\odot}$.
We adopt the Solar photospheric abundances provided by \citet{asplund09}.
By convention, abundances or ratios denoted with the ionization state indicate the total elemental abundance derived from transitions of that particular ionization state after Saha ionization corrections have been applied. If there is no ionization state indicated, the abundance also represents the total elemental abundance after applying the ionization correction.

We derive abundances for all elements for which we have EW measurements, based on lines of O~\textsc{i}, Na~\textsc{i}, Mg~\textsc{i}, Si~\textsc{i}, K~\textsc{i}, Ca~\textsc{i}, Ti~\textsc{i}, Ti~\textsc{ii}, Cr~\textsc{i}, Cr~\textsc{ii}, Fe~\textsc{i}, Fe~\textsc{ii}, Ni~\textsc{i}, and Zn~\textsc{i}.
These lines are listed in Table~\ref{abundtab}.
These EW measurements were also made by I.U.R., but they were not published previously as part of the \citet{roederer14c} study.

\subsection{LTE Calculations}
\label{abundcalc}

We derive abundances following the same procedure described in Section~\ref{microturbulent}.
MOOG's ``abfind'' driver calculates the abundance required to yield the measured EW of each line of each species. Table~\ref{abundtab} lists the species, wavelength ($\lambda$), EW, and abundance for each line in each star. The individual abundance ($\log\epsilon$(X)) values for each line of a given element in a given star were combined to give a single inverse-weighted average $\log\epsilon$(X) value and its associated uncertainty as described in Section~\ref{unc}. These values are listed in Table~\ref{ltetab}.

\begin{deluxetable*}{lcccccccccccccc}
\tablecaption{Derived LTE Abundances
\label{ltetab}}
\tablewidth{0pt}
\tabletypesize{\scriptsize}
\tablehead{
\colhead{Star} &
\colhead{$N_{\rm lines}$} &
\colhead{$\log\epsilon$(Fe~\textsc{i})} &
\colhead{[Fe~\textsc{i}/H]} &
\colhead{$\sigma$} &
\colhead{$N_{\rm lines}$} &
\colhead{$\log\epsilon$(Fe~\textsc{ii})} &
\colhead{[Fe~\textsc{ii}/H]} &
\colhead{$\sigma$} &
\colhead{$N_{\rm lines}$} &
\colhead{$\log\epsilon$(O~\textsc{i})} &
\colhead{[O~\textsc{i}/Fe]} &
\colhead{$\sigma$} &
\colhead{...} \\
\colhead{} &
\colhead{Fe~\textsc{i}} &
\colhead{LTE} &
\colhead{LTE} &
\colhead{} &
\colhead{Fe~\textsc{ii}} &
\colhead{LTE} &
\colhead{LTE} &
\colhead{} &
\colhead{O~\textsc{i}} &
\colhead{LTE} &
\colhead{LTE} &
\colhead{} &
\colhead{}
}
\startdata
\mbox{BD+10\degree2495} & 81 & 5.54 & $-$1.96 & 0.05 & 8 & 5.59 & $-$1.91 & 0.15 & 3 & 7.36 & 0.55 & 0.10 & ...\\
\mbox{BD+19\degree1185A} & 54 & 6.34 & $-$1.16 & 0.05 & 9 & 6.36 & $-$1.14 & 0.15 & 3 & 8.16 & 0.61 & 0.10 & ...\\
\mbox{BD+24\degree1676} & 85 & 5.10 & $-$2.40 & 0.05 & 11 & 5.14 & $-$2.36 & 0.17 & 2 & 6.99 & 0.61 & 0.12 & ...\\
\mbox{BD+26\degree3578} & 76 & 5.35 & $-$2.15 & 0.05 & 10 & 5.32 & $-$2.18 & 0.17 & 3 & 6.99 & 0.36 & 0.11 & ...\\
\mbox{BD+29\degree2356} & 87 & 5.97 & $-$1.53 & 0.05 & 8 & 5.79 & $-$1.71 & 0.18 & 1 & 7.96 & 0.71 & 0.16 & ...\\
\enddata      
\tablecomments{%
Abundances derived from O~\textsc{i}, Na~\textsc{i}, Mg~\textsc{i}, Si~\textsc{i}, K~\textsc{i}, Fe~\textsc{i}, and Fe~\textsc{ii} lines have been corrected for NLTE. Their LTE abundances are presented in Table~\ref{ltetab} for reference, but we recommend adopting the NLTE-corrected abundances presented in Table~\ref{nltetab}. The complete version of Table~\ref{ltetab} is available in the online edition of the journal. A short version is included here to demonstrate its form and content.
}
\end{deluxetable*}

\subsection{Non LTE Corrections}
\label{nlte}

We interpolate line-by-line NLTE corrections from the INSPECT database, version 1.0\footnote{\href{www.inspect-stars.net}{www.inspect-stars.net}} for O \citep{amarsi2015}, Na \citep{lind11}, and Fe~\textsc{i} \citep{bergemann2012,lind12}.

We interpolate NLTE corrections from the MPIA NLTE database \citep{NLTE_MPIA}\footnote{\href{https://nlte.mpia.de/gui-siuAC_secE.php}{https://nlte.mpia.de/gui-siuAC\_secE.php}} for Mg \citep{bergemann2015}, Si \citep{bergemann2013}, and Fe~\textsc{ii} \citep{bergemann2012}. We developed a new code\footnote{\href{https://github.com/sanilmittal/nlte-corr}{https://github.com/sanilmittal/nlte-corr}} to interpolate NLTE corrections from these online databases in batch mode, so that we can query and return multiple NLTE corrections simultaneously, rather than one line or one star at a time. For K, we interpolate NLTE corrections from the tables presented by \cite{takeda02} (private communication, 2007). Whenever the stellar parameters or metallicities of the star of interest extend beyond the grid of NLTE models, we adopt the NLTE correction for the closest model available.

The individual NLTE-corrected abundance values ($\log\epsilon_{N}$(X)) for all lines for which NLTE corrections of a given element were computed in a given star were combined to give a single inverse-weighted average $\log\epsilon_N$(X) value and its associated uncertainty as described in Section~\ref{unc}. These values are listed in Table~\ref{nltetab}.

\begin{deluxetable*}{lcccccccccccccc}
\tablecaption{Derived Non-LTE Abundances
\label{nltetab}}
\tablewidth{0pt}
\tabletypesize{\scriptsize}
\tablehead{
\colhead{Star} &
\colhead{$N_{\rm lines}$} &
\colhead{$\log\epsilon$(Fe~\textsc{i})} &
\colhead{[Fe~\textsc{i}/H]} &
\colhead{$\sigma$} &
\colhead{$N_{\rm lines}$} &
\colhead{$\log\epsilon$(Fe~\textsc{ii})} &
\colhead{[Fe~\textsc{ii}/H]} &
\colhead{$\sigma$} &
\colhead{$N_{\rm lines}$} &
\colhead{$\log\epsilon$(O~\textsc{i})} &
\colhead{[O~\textsc{i}/Fe]} &
\colhead{$\sigma$} &
\colhead{...} \\
\colhead{} &
\colhead{Fe~\textsc{i}} &
\colhead{NLTE} &
\colhead{NLTE} &
\colhead{} &
\colhead{Fe~\textsc{ii}} &
\colhead{NLTE} &
\colhead{NLTE} &
\colhead{} &
\colhead{O~\textsc{i}} &
\colhead{NLTE} &
\colhead{NLTE} &
\colhead{} &
\colhead{}
}
\startdata
\mbox{BD+10\degree2495} & 19 & 5.62 & $-$1.88 & 0.07 & 8 & 5.59 & $-$1.91 & 0.16 & 3 & 7.23 & 0.42 & 0.14 & ...\\
\mbox{BD+19\degree1185A} & 14 & 6.36 & $-$1.14 & 0.08 & 9 & 6.33 & $-$1.17 & 0.16 & 3 & 8.07 & 0.52 & 0.15 & ...\\
\mbox{BD+24\degree1676} & 15 & 5.19 & $-$2.31 & 0.10 & 11 & 5.14 & $-$2.36 & 0.18 & 2 & 6.86 & 0.48 & 0.18 & ...\\
\mbox{BD+26\degree3578} & 13 & 5.44 & $-$2.06 & 0.09 & 10 & 5.32 & $-$2.18 & 0.18 & 3 & 6.86 & 0.23 & 0.15 & ...\\
\mbox{BD+29\degree2356} & 24 & 6.06 & $-$1.44 & 0.07 & 8 & 5.79 & $-$1.71 & 0.19 & 1 & 7.81 & 0.56 & 0.24 & ...\\
\enddata      
\tablecomments{%
The number of lines listed in Table~\ref{nltetab} refers to the number of lines for with NLTE corrections were calculated. These corrections have been added to the LTE abundances for individual lines to recompute the NLTE abundances listed here using the procedure explained in Section~\ref{unc}. For O~\textsc{i}, Na~\textsc{i}, Mg~\textsc{i}, Si~\textsc{i}, K~\textsc{i}, Fe~\textsc{i}, and Fe~\textsc{ii}, we recommend adopting the NLTE-corrected abundances presented in Table~\ref{nltetab}.
The complete version of Table~\ref{nltetab} is available in the online edition of the journal. A short version is included here to demonstrate its form and content.
}
\end{deluxetable*}

\subsection{Abundance and Uncertainty calculations}
\label{unc}

We estimate uncertainties in the $\log\epsilon$ abundances
and [X/Fe] ratios by combining the statistical uncertainty (standard deviation of the $\log\epsilon$(X) values obtained from different lines for a given element in a given star) with the uncertainties that result from the uncertainties in the stellar parameters.
We calculate the latter uncertainties using the method described in appendix B of \citet{ji2020} using their equations B10--B13.

We calculate the individual line uncertainties ($e_i$) by adding the uncertainties in equivalent widths ($\sigma_{\rm EW}$) and a factor of 0.02~dex (to account for the typical uncertainties in the \loggf\ values) in quadrature:
\begin{equation}
    e_i = \sqrt{\sigma_{\rm EW}^2 + 0.02^2}.
\end{equation}
For NLTE corrections, we add an additional uncertainty of 0.05~dex in quadrature:
\begin{equation}
    e_i = \sqrt{\sigma_{\rm EW}^2 + 0.02^2 + 0.05^2}.
\end{equation}
Since we do not measure the equivalent width uncertainties directly, we estimate them based on the signal-to-noise ratio (SNR) at the wavelength for each line linearly scaled to the SNR at 5200 \AA\ for the star (table 3 of \citealt{roederer14c}), by defining a simple relation, 
\begin{equation}
    \sigma_{\rm EW} = {\rm EW}_{c}(R)S.
\end{equation}
We scale the changes in abundance for a change in EW of 1~m\AA\ (${\rm EW}_c(R)$) as a function of the line strength ($R$) using the scale factor $S$ defined below. 
$R$ is the reduced equivalent 
width, commonly defined in the field as
\begin{equation}
    R = \log_{10}(\frac{{\rm EW}}{\lambda}).
\end{equation}
These values are obtained from the coefficients for a polynomial fit of order 5 to the data in Figure~19 in \cite{roederer14c}.
$S$ is the scale factor for individual lines with wavelength $\lambda$, which we estimate empirically using the function
\begin{equation}
    S = \frac{10^{17} \lambda^{-4}}{\text{SNR}}.
\end{equation}
For example, the estimated $\sigma_{\rm EW}$ value for a line at $\lambda = 4000$~\AA\ with EW $= 50$~m\AA\ ($R = -4.90$) in a star with SNR $= 100$ at 5000~\AA\ 
($S = 3.91$) would be \mbox{(0.02)(3.91)} = 0.08~dex.

We construct the $\delta$ matrix as defined in appendix B of \cite{ji2020} by obtaining the individual stellar parameter error ($\delta\theta_k$) to populate the matrix, where $\theta_k$ refers to the star's measured stellar parameters with $k$=1--4 ($\theta_1$=\teff, $\theta_2$=\logg, $\theta_3$=\vt, and $\theta_4$=[M/H]). We first find the coefficients ($\theta_{k,c}$) for each stellar parameter by fitting an order 5 polynomial to the data presented in figures~16--18 in \cite{roederer14c}. We adopt an uncertainty of 0.00~dex for the uncertainty in $\theta_{\feh}$, because the abundances are insensitive to this parameter. These values provide us with the changes in abundance for a change in a stellar parameter ($\theta_k$) obtained for a given line strength ($R$), which we scale by our absolute uncertainty for $\theta_k$ ($\sigma_{\theta_k}$) from Section~\ref{params}. These fit coefficients $\theta_{k,c}$ differ for stars in different evolutionary states, so we approximate the evolutionary states for our data using Figure~\ref{tefflogg} and roughly divide it as follows.  Main sequence (MS) stars are defined to have $\logg > 4.0$, subgiants (SG) are defined to have $3.0 \leq \logg  \leq 4.0$, horizontal branch (HB) stars are defined to have \logg $<$ 3.0 and \teff $>$ 5400~K, and all remaining stars are considered to be red giants (RG).~
One star, \mbox{CS~29504--004}, may be a blue straggler (BS) warmer than the main sequence turnoff, and we classify it separately.

We combine these input data to construct the individual $\delta$ matrix elements using:
\begin{equation}
    \delta\theta_k = \frac{\sigma_{\theta_k}}{c}\theta_{k,c}(R) 
\end{equation}
where $c$ is a constant that scales the values and their units correctly as given on the y-axes of Figures~16--18 in \cite{roederer14c}. For example, when calculating the uncertainty contribution by \teff, $c$ is 100~K. For \logg and \vt, it is 0.4~dex and 0.4~\kmsec, respectively. We enforce an artificial minium uncertainty of 0.01~dex for all of the individual values ($e_i, \delta\theta_k$) to guard against unreasonably small values for the uncertainty. Finally, we input the constructed $\delta$ matrix and $e_i$ values in equations B10--B13 in appendix B of \citet{ji2020} to obtain the abundances ($\hat{x}$) and the associated uncertainties ($\sqrt{\text{Var}(\hat{x})}$).

\section{Results}
\label{results}
Our goal in this study is to rederive stellar parameters and abundances for the sample of stars presented by \citet{roederer14c} using a photometric \teff\ scale and \logg\ values calculated using distances or parallaxes. 
%
%
We present the final model atmosphere parameters in Table~\ref{phottab}.
Our derived line-by-line LTE abundances, including NLTE corrections when available, are presented in Table~\ref{abundtab}.

\begin{deluxetable*}{cccccccccccccc}
\tablecaption{Comparison of Model Atmosphere Parameters
\label{comptab}}
\tablewidth{0pt}
\tabletypesize{\small}
\tablehead{
\colhead{Class\tablenotemark{a}} &
\colhead{Number} &
\colhead{Avg.\ \teff} &
\colhead{$\Delta$\teff} &
\colhead{$\sigma$\teff} &
\colhead{Avg. \logg} &
\colhead{$\Delta$\logg}&
\colhead{$\sigma$\logg} &
\colhead{Avg. [M/H]} & 
\colhead{$\Delta$[M/H]} &
\colhead{$\sigma$[M/H]} &
\colhead{Avg. \vt} & 
\colhead{$\Delta$ \vt} &
\colhead{$\sigma$ \vt} \\
\colhead{} &
\colhead{} &
\colhead{(K)} &
\colhead{(K)} &
\colhead{(K)} &
\colhead{[cgs]} &
\colhead{[cgs]}&
\colhead{[cgs]} &
\colhead{} & 
\colhead{(dex)} &
\colhead{(dex)} &
\colhead{(\kmsec)} & 
\colhead{(\kmsec)} &
\colhead{(\kmsec)} 
}
\startdata
MS & 143 & 6240 & $+$362 & 12 & 4.31 & $+$0.57 & 0.03 & $-$2.24 & $+$0.26 & 0.01 & 1.37 & $+$0.07 & 0.02 \\
SG & 53 & 5965 & $+$319 & 26 & 3.67 & $+$0.40 & 0.05 & $-$2.39 & $+$0.18 & 0.02 & 1.48 & $+$0.16 & 0.02  \\
RG & 79 & 5079 & $+$311 & 10 & 2.25 & $+$0.86 & 0.04 & $-$2.72 & $+$0.34 & 0.02 & 1.71 & $+$0.12 & 0.01 \\
HB & 35 & 5808 & $+$98 & 25 & 2.52 & $+$0.78 & 0.05 & $-$2.32 & $+$0.25 & 0.02 & 2.56 & $+$0.08 & 0.01 \\
BS & 1 & 7451 & $-$49 & \ldots & 3.11 & $+$0.61 & \ldots & $-$2.20 & $+$0.15 & \ldots & 2.02 & $+$0.07 & \ldots 
\enddata      
\tablenotetext{a}{The classification scheme is defined as follows: BS, blue straggler-like stars warmer than the main sequence turnoff; HB, stars on the horizontal branch; MS, stars on the main sequence; RG, stars on the red giant branch; SG, stars on the subgiant branch.}

\tablecomments{``Avg.'' refers to the mean of the updated stellar parameter of all stars in the respective evolutionary class. ``$\Delta$'' refers to the mean difference in the respective stellar parameter for each star in an evolutionary class. ``$\sigma$'' refers to the standard deviation of the difference of the stellar parameter for stars in an evolutionary class, divided by the square root of the total number of stars in that evolutionary class.}

\end{deluxetable*}

\begin{figure}
\includegraphics[angle=0,width=3.35in]{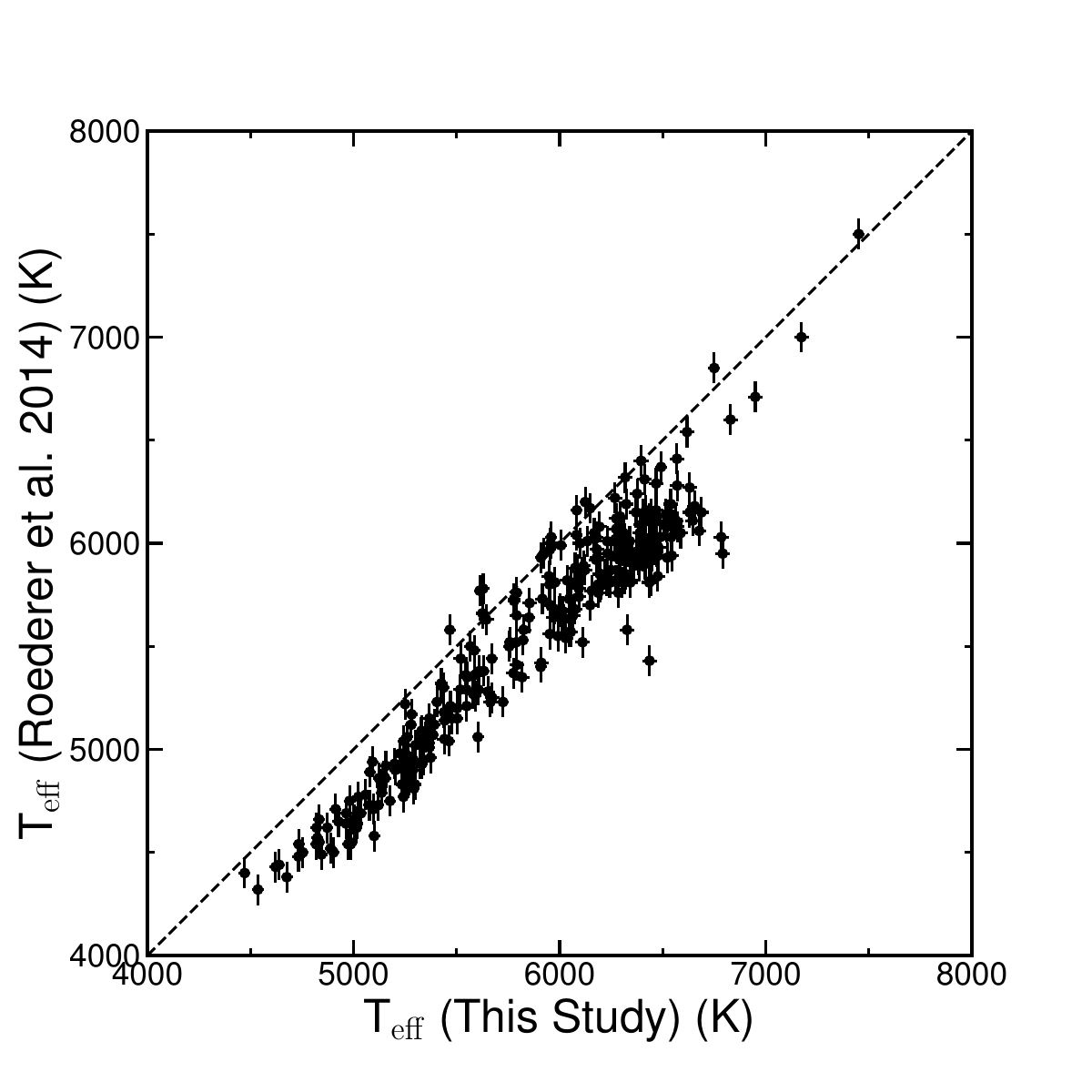}
\caption{
\label{teffplot}
Comparison of \teff values found in this study and \cite{roederer14c}. The new values are 310~$\pm$~164~K warmer on average. The dashed line indicates a 1:1 ratio. Only statistical uncertainties are shown. The size of the error bars are comparable to the size of the markers, making it difficult to see them.}
\end{figure}

We first compare our results with those from \citet{roederer14c}. These comparisons are listed in Table~\ref{comptab}, with differences calculated in the sense of ``new'' minus ``old.'' The new \teff values are warmer in all evolutionary tracks, as shown in Figure~\ref{teffplot}. It is unclear why the mean \teff differences between warmer (SG) stars and cooler (RG) stars are similar. As shown in Figure~\ref{teffplot}, the RG stars exhibit a consistently warmer difference, whereas the SG stars exhibit a wider distribution of differences. We find no obvious errors in our code or our implementation of the color-\teff relations. \citeauthor{roederer14c}\ adopted mixed methods to calculate stellar parameters for the stars in different evolutionary states.  For the SG and RG stars, that study calculated \logg values from isochrones; for other stars, that study derived \logg values from the Fe ionization balance. That study then iterated to converge on other methods of determining \teff, \logg, and \vt. That approach was reasonable for a large, inhomogeneous sample of stars, and before Gaia data were available.  Nevertheless this iterative process may have led to different convergences in the final stellar parameters, which may not have been as homogeneous as one would have preferred. Our current methods avoid these issues, and our calculated stellar parameters should be more reliable and homogeneous.

\begin{figure}
\includegraphics[angle=0,width=3.35in]{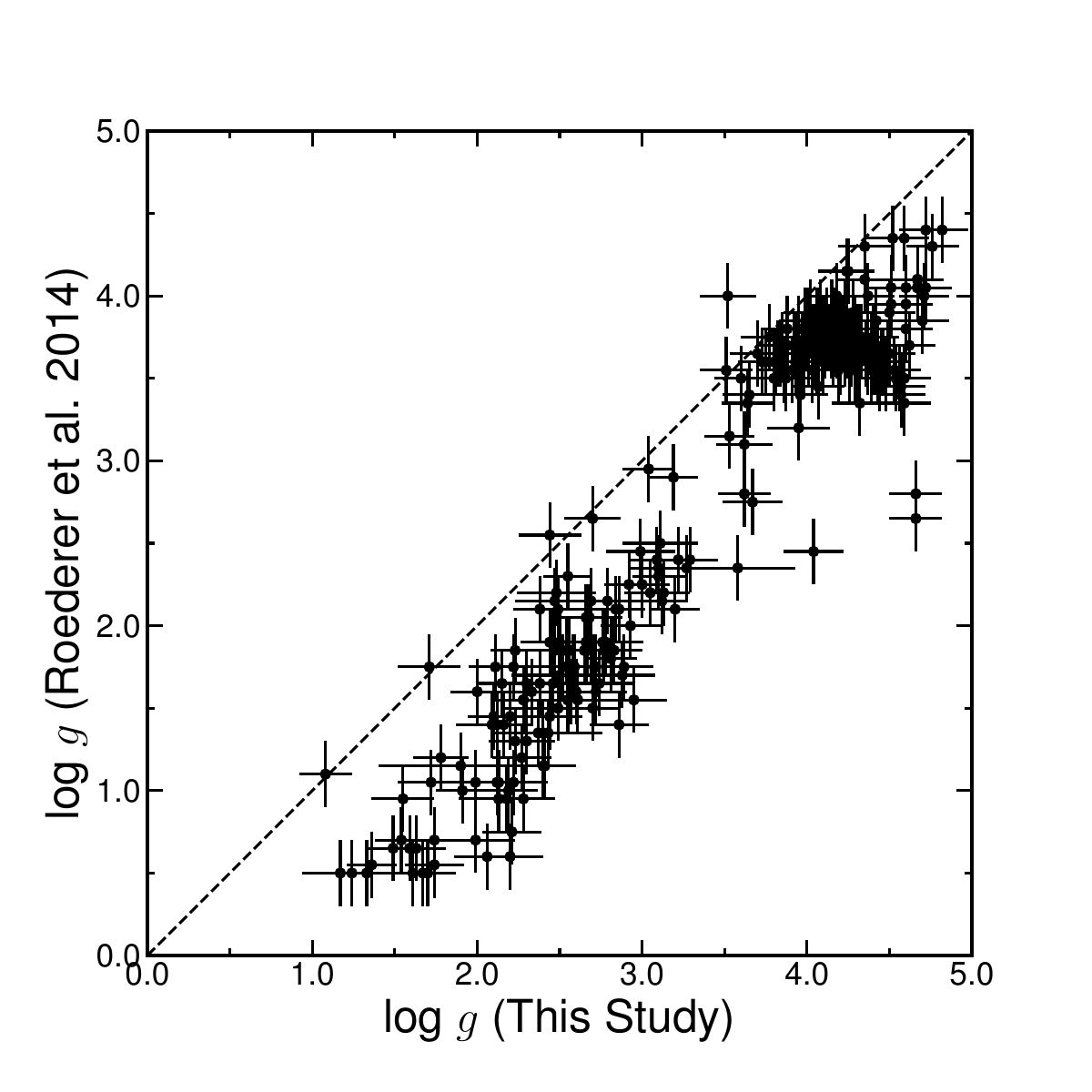}
\caption{
\label{loggplot}
Comparison of \logg values found in this study and \cite{roederer14c}. The new values are 0.64~$\pm$~0.35 higher on average. The dashed line indicates a 1:1 ratio.
Stars that lie along the 1:1 line are stars for which precise parallax values were available to \citeauthor{roederer14c} }
\end{figure}

\begin{figure}
\includegraphics[angle=0,width=3.35in]{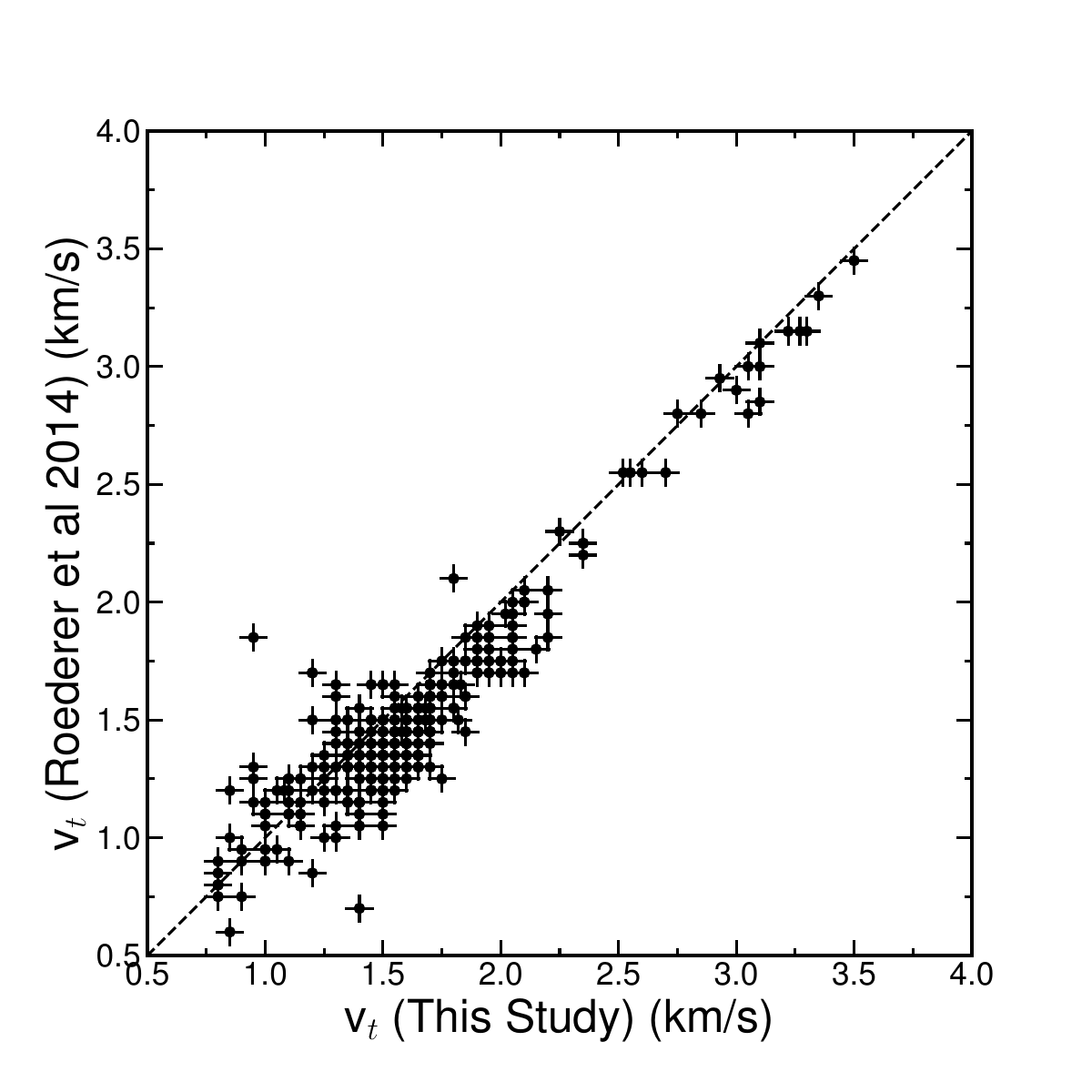}
\caption{
\label{vtplot}
Comparison of microturbulence velocity \vt\ values found in this study and \cite{roederer14c}. The new values are 0.06~$\pm$~0.19~\kmsec\ higher on average. The dashed line indicates a 1:1 ratio.}
\end{figure}

\begin{figure}
\includegraphics[angle=0,width=3.35in]{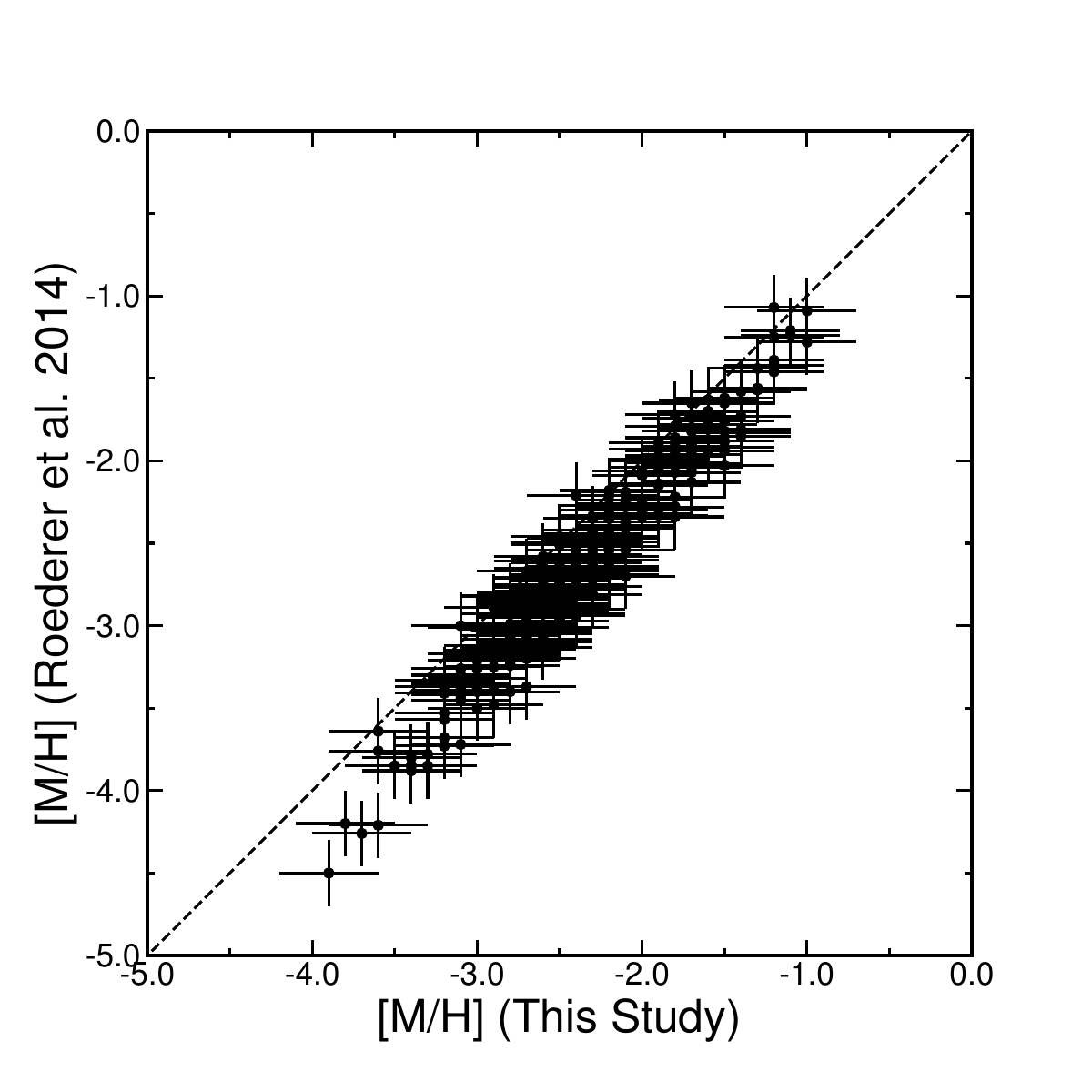}
\caption{
\label{mhplot}
Comparison of model metallicity [M/H] values found in this study and \cite{roederer14c}. The new values are on average 0.27~$\pm$~0.14~dex higher. The dashed line indicates a 1:1 ratio.
}
\end{figure}

The new \logg values are higher (Figure \ref{loggplot}), the new \vt\ values are higher (Figure \ref{vtplot}), and the new model metallicities are higher (Figure \ref{mhplot}).  We also calculate that the new [Fe/H] values are higher than those presented by \citeauthor{roederer14c}\ on average. The average differences for individual evolutionary classes are listed in Table~\ref{comptab}.

\begin{figure}
\includegraphics[angle=0,width=3.35in]{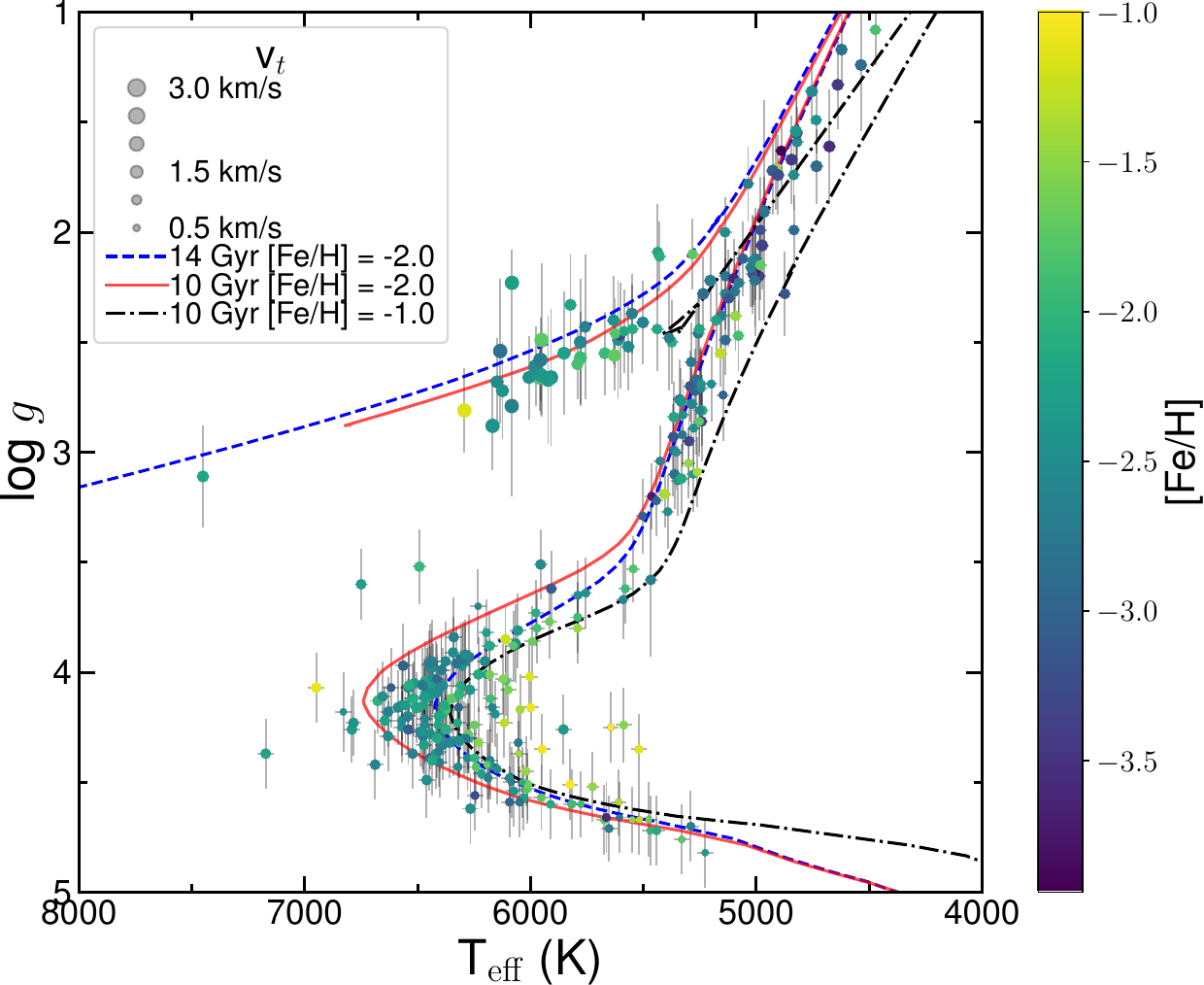}
\caption{
\label{tefflogg}
Plot of \teff and \logg values, color-coded by [M/H] values. The size of the marker is given by the \vt\ value. We also plot Padova Isochrones with $\feh$ = $-2.0$; the red line for an age of 10 Gyr, and the dashed blue line for an age of 14 Gyr.
}
\end{figure}

We show the \teff and \logg values of the stars in our analysis in Figure \ref{tefflogg}. For comparison, we plot Padova isochrones\footnote{\href{http://stev.oapd.inaf.it/cmd}{http://stev.oapd.inaf.it/cmd}} (\citealp{bressan12}; \citealp{chen2014,chen2015}; \citealp{tang2014}; \citealp{marigo2017}; \citealp{pastorelli2019,pastorelli2020}) to demonstrate that the new stellar parameters trace out the expected regions of the \teff-\logg diagram for old, metal-poor stellar populations.

\subsection{Iron Abundances}

\begin{figure}
\includegraphics[angle=0,width=3.35in]{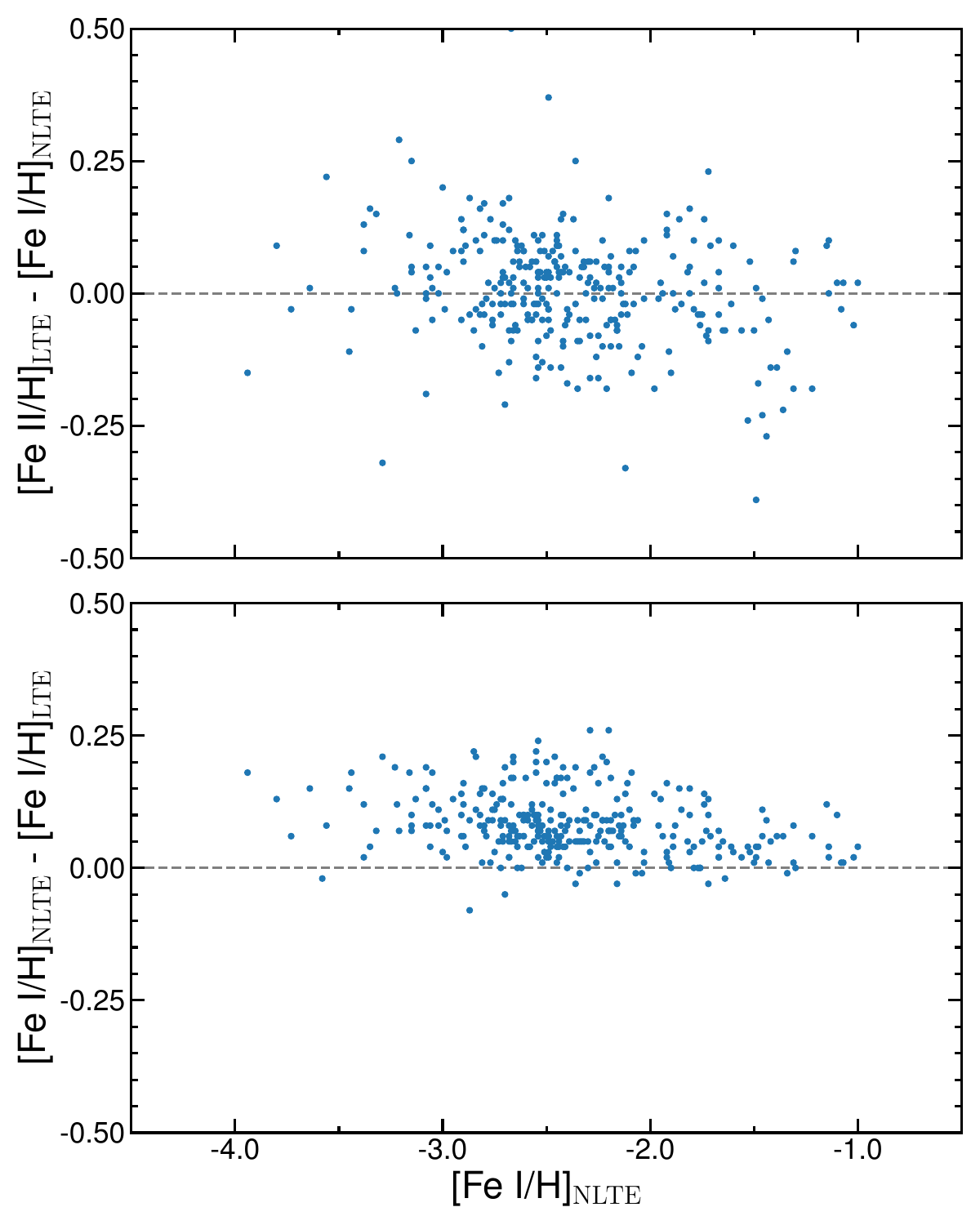}
\caption{
\label{deltafeh}
Comparison of NLTE and LTE abundances for different Fe ionization states. The top panel shows the difference between the LTE [Fe~\textsc{ii}/H] and NLTE-corrected [Fe~\textsc{i}/H] values. The bottom panel shows the amount of NLTE corrections applied to [Fe~\textsc{i}/H]. In both panels, the dashed line indicates a difference of 0.0. In the bottom panel, the stars with differences $<$ 0 are MS stars with \logg $>$ 4, where NLTE effects are expected to be minimal (e.g., \citealt{lind12}).
}
\end{figure}

The top panel of Figure~\ref{deltafeh} shows that the NLTE-corrected abundances based on Fe~\textsc{i} lines are in agreement with the abundances based on Fe~\textsc{ii} lines.
Although the Fe~\textsc{ii} lines are expected to be formed near LTE conditions, many more Fe~\textsc{i} lines are present in these spectra than Fe~\textsc{ii} lines, so the abundances derived from Fe~\textsc{i} lines are typically more precise.
We thus adopt the NLTE-corrected abundances based on Fe~\textsc{i} lines as the best measure of the iron abundances in this sample of stars.

The bottom panel of Figure~\ref{deltafeh} shows the difference between the NLTE and LTE abundances derived from Fe~\textsc{i} lines. 
The sign and magnitude of the differences here reflect the overionization of neutral iron in these metal-poor stars.

\subsection{Other Elements}

\begin{figure*}[h!]
\centering
\includegraphics[angle=0, width=6.7in]{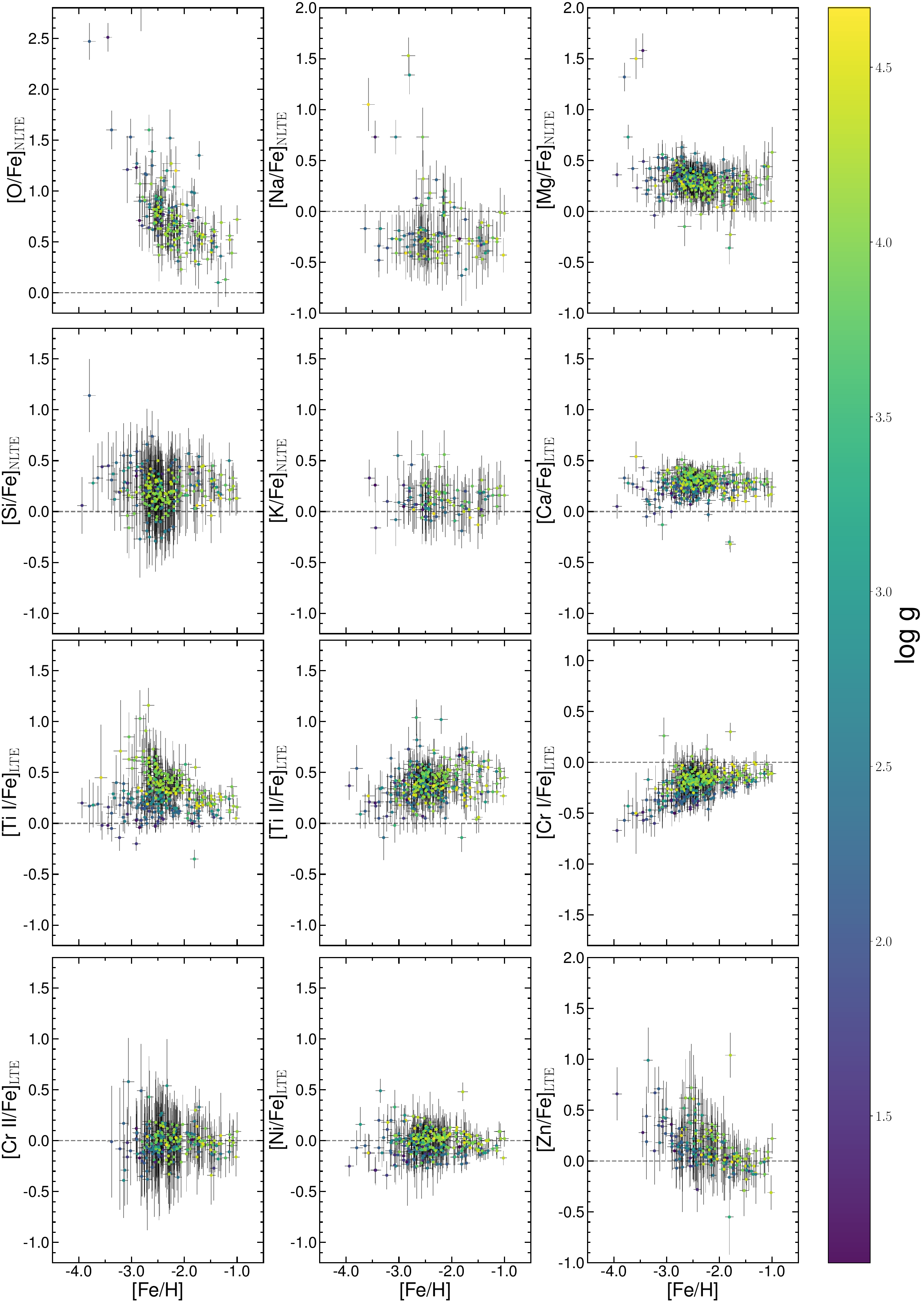}
\caption{
\label{elements}
Elemental abundance ratios as a function of [Fe~\textsc{i}/H]. The dashed line indicates the Solar ratio in all of the panels.}
\end{figure*}

We present our derived abundances for other elements in Table~\ref{ltetab}. Table~\ref{nltetab} presents the NLTE corrected abundances for O, Na, Mg, Si, K, and Fe. 
These results are shown in Figure~\ref{elements}, where we plot the [X/Fe] ratio, for each element X, as a function of [Fe/H]. The y-axis of each subplot spans the same number of dex, but they are centered at different values. This display choice helps to illuminate the star-to-star dispersion for each ratio and to compare these ratios in metal-poor stars to the Solar ratios.

\section{Discussion}
\label{discussion}

\begin{figure*}[h!]
    \centering
    \includegraphics[angle=0, width=6.7in]{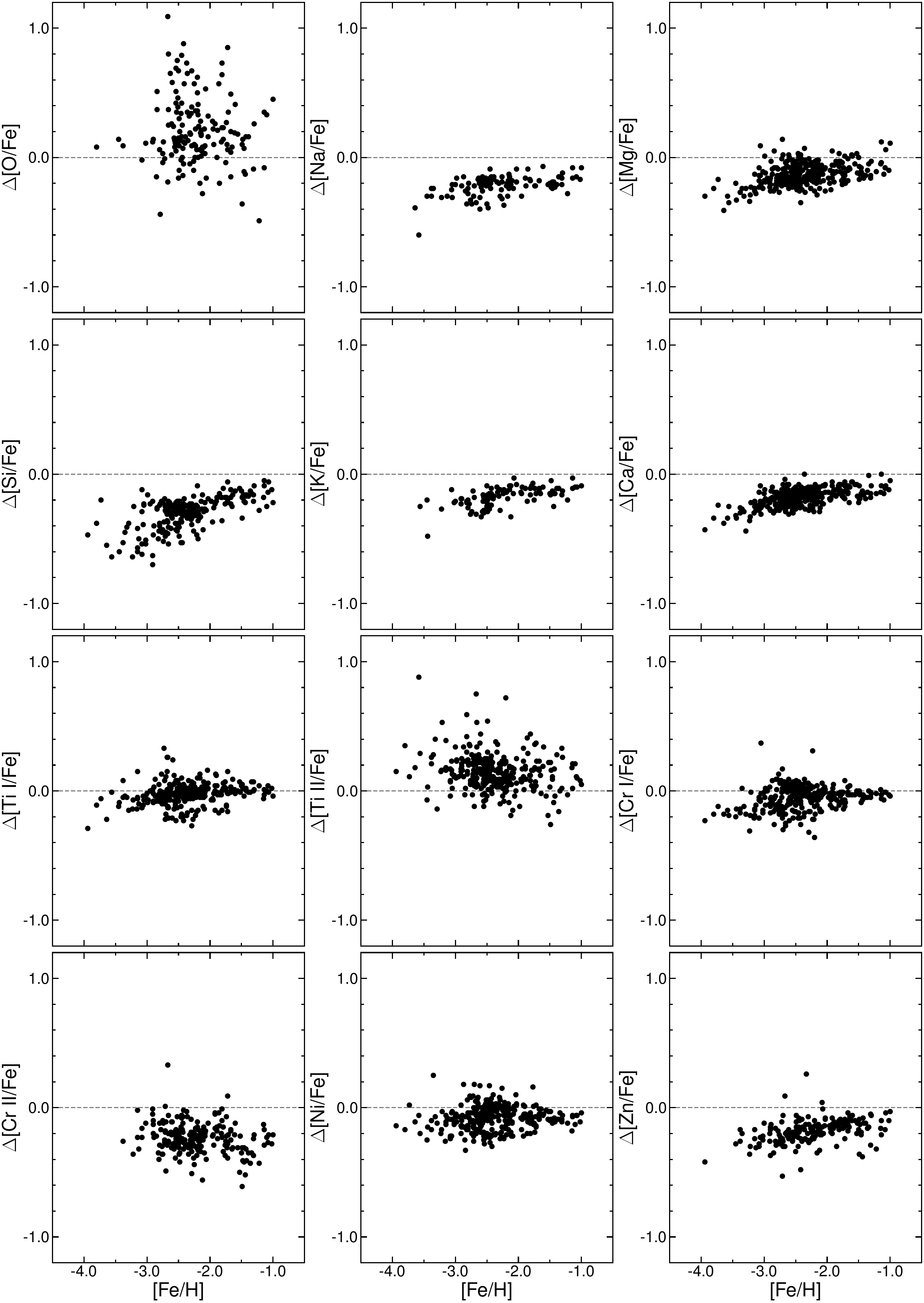}
    \caption{\label{fig:iur}
    Differences between our abundances and abundances published by \cite{roederer14c}. The dashed line indicates the solar ratio in all of the panels. $\Delta$ refers to the updated abundance minus previously published abundance. NLTE corrections have been applied to the O, Na, Mg, Si, K, and Fe abundances.}
    
\end{figure*}

We compare our derived [X/Fe] ratios, where X is any given element, with the abundances published by \cite{roederer14c} in Figure~\ref{fig:iur}. The differences are plotted as a function of $\feh$ values for the respective stars adopted by this study. Our 
NLTE-corrected values are used for O, Na, Mg, Si, and K. The differences in the [X/Fe] ratios are generally the largest for the stars with the lowest metallicities and/or coolest temperatures, where the stellar parameters exhibited the largest differences. The change in [X/Fe] ratios from \citeauthor{roederer14c}\ to this study is typically small, because both elements respond similarly to changes in the model atmosphere parameters.

The [O/Fe] ratios are considerably larger in our study for two reasons.  First, we have adopted NLTE corrections from a different study than was used by \citet{roederer14c}. Our adopted corrections, while still negative, are typically smaller in magnitude than those adopted previously. Second, and more substantially, that study applied an additional $-0.50$~dex correction to the [O/Fe] ratios to account for the typical difference between the high-excitation near-infrared O~\textsc{i} triplet and the forbidden [O~\textsc{i}] line at 6300~\AA, which is formed under conditions very nearly in LTE \citep{kiselman01}. Our study avoids all such empirical corrections. When these two effects are accounted for, the [O/Fe] ratios between the two studies are approximately reconciled.

\begin{figure*}[h!]
    \centering
    \includegraphics[angle=0, width=6.7in]{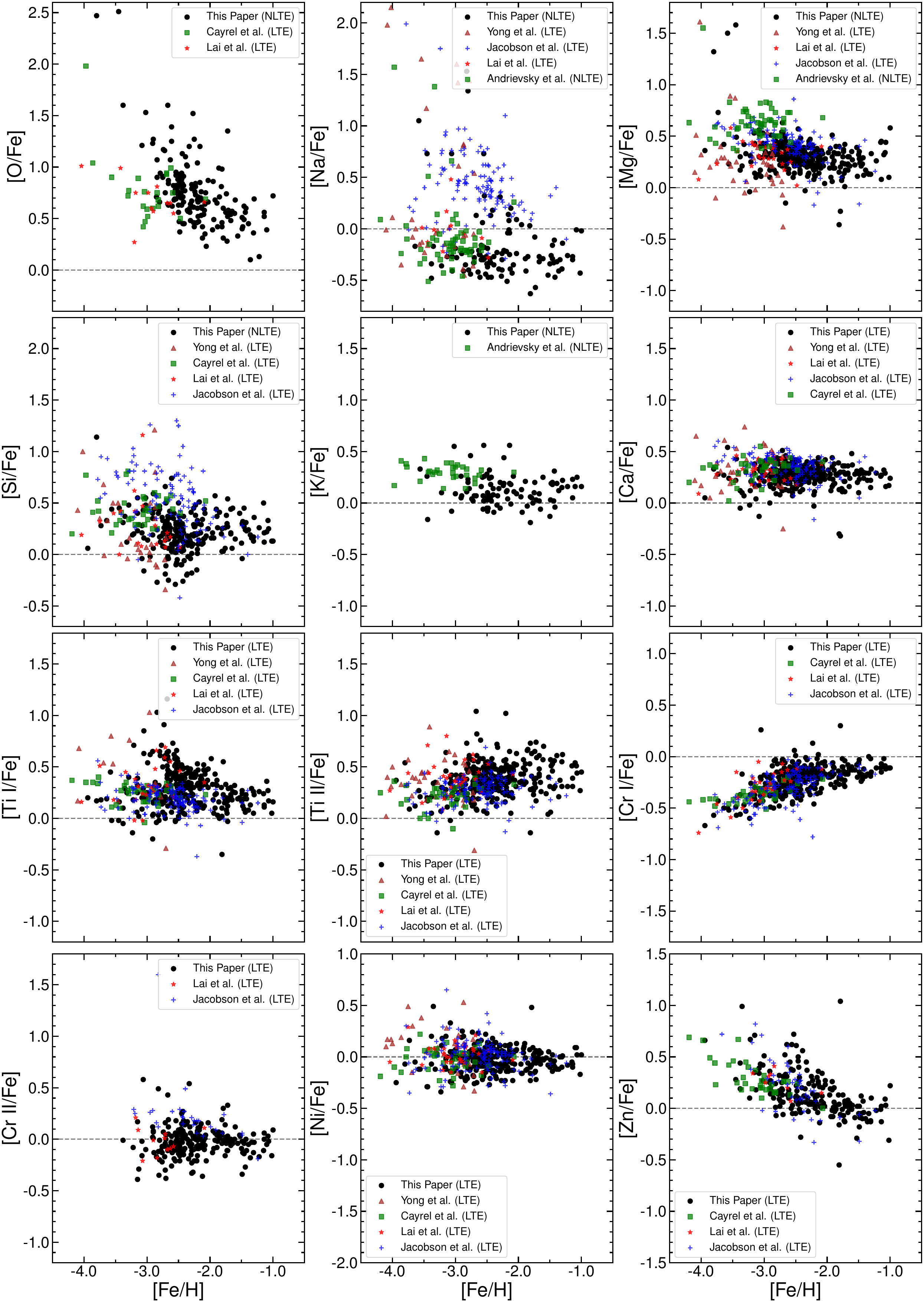}
    \caption{\label{fig:feh-comparison}
    Comparison of our abundances with other metal-poor star samples. The references for the symbols are given in the panel legends and the main body of the text. The dashed line indicates the solar ratio in all of the panels.}
\end{figure*}

We compare our derived [X/Fe] ratios with those derived by previous studies in Figure~\ref{fig:feh-comparison}. We perform a straight comparison of the results without accounting for the differences in, e.g., the number of lines or the set of lines used for a given element. We use the following studies in our comparison: \cite{cayrel04}, \cite{lai08}, \cite{yong13a}, and \cite{jacobson15smss}. We update the Na, Mg, and K abundances from \citeauthor{cayrel04} with the NLTE-corrected ones in \cite{andrievsky07} and \cite{andrievsky2010}.

For most of the element ratios, the agreement is excellent. The range of derived [X/Fe] ratios follows the same trends in the regions where the [Fe/H] values overlap. The [Na/Fe] ratios exhibit noticeable differences only when compared to \cite{jacobson15smss}, who did not publish NLTE-corrected Na abundances. NLTE-corrected abundances are shown in their figure~8, and they are in the same range as our values. \citeauthor{jacobson15smss}\ state in their section~4.3 that their NLTE corrections are large ($-0.7$ to $-1.0$~dex), affirming that the difference shown in our Figure~\ref{fig:feh-comparison} are due to NLTE corrections in Na and Fe abundances.

The previously observed offsets of the \cite{roederer14c} study in comparison to other metal-poor star samples have now been resolved by our use of photometric \teff values and \logg values ultimately derived from Gaia parallaxes.

\section{Conclusions}
\label{conclusions}

The sample of 313 metal-poor stars first analyzed by \citet{roederer14c} remains in common use today. The present study consistently used the latest color-\teff relations calibrated by \cite{mucciarelli2021} for Gaia DR3 colors to rederive \teff values, compared with the original \citeauthor{roederer14c} study. These are then used to calculate \logg values. The new stellar parameters are then used to generate 1-D MARCS model atmospheres, which are input to MOOG to rederive \vt\ values and chemical abundances based on our high resolution optical spectroscopy. These results deviate from the original \citeauthor{roederer14c} work, which used primarily spectroscopic methods to derive their stellar parameters.

The new \teff values are 310~$\pm$~164~K warmer on average. The new \logg values are 0.64~$\pm$~0.35 higher on average. The new \vt\ values are 0.06~$\pm$~0.19~\kmsec\ higher on average. The different evolutionary classes of stars exhibit slightly different behaviors, as listed in Table~\ref{comptab}. We adopt NLTE-corrected abundances derived from Fe~\textsc{i} lines as the new $\feh$ values, which are higher by 0.26~$\pm$~0.15~dex, on average, in comparison to the previously adopted $\feh$ values derived from Fe~\textsc{ii} lines by \cite{roederer14c}. Our sample contains 6 stars with $\feh < −3.5$, 24 stars with $\feh < −3.0$, and 121 stars with $\feh < −2.5$.

We perform a standard LTE abundance analysis to derive abundances for 14 species of 11 elements (O~\textsc{i}, Na~\textsc{i}, Mg~\textsc{i}, Si~\textsc{i}, K~\textsc{i}, Ca~\textsc{i}, Ti~\textsc{i}, Ti~\textsc{ii}, Cr~\textsc{i}, Cr~\textsc{ii}, Fe~\textsc{i}, Fe~\textsc{ii}, Ni~\textsc{i}, and Zn~\textsc{i}). We apply NLTE corrections to abundances of O, Na, Mg, Si, K, and Fe. 

These stellar parameters and abundances can now be used as a more direct comparison sample for other metal-poor star surveys. We have resolved the discrepancy in [X/Fe] ratios when compared with other stellar samples, as noted by \cite{roederer14c}, by using photometric temperatures and \logg values based on parallaxes or distances. Our elemental abundances match the results by other large studies performed for metal-poor stars in the Milky Way. Detailed abundances for elements requiring spectrum synthesis, such as Al, V, Mn, Cu, Y, Zr, Eu, etc., are beyond the scope of this paper and will be presented separately.

\section{Acknowledgements}

We acknowledge financial support from
grants AST-1815403, AST-2205847,
and PHY~14-30152 (Physics Frontier Center/JINA-CEE)
awarded by the U.S.\ National Science Foundation (NSF).
This research has made use of NASA's
Astrophysics Data System Bibliographic Services;
the arXiv pre-print server operated by Cornell University;
the SIMBAD and VizieR databases hosted by the
Strasbourg Astronomical Data Center;
the Atomic Spectra Database hosted by
the National Institute of Standards and Technology;
and the cross-match service provided by CDS, Strasbourg.
This work has made use of data from the European Space Agency (ESA) mission
{\it Gaia} (\url{https://www.cosmos.esa.int/gaia}), processed by the {\it Gaia}
Data Processing and Analysis Consortium (DPAC,
\url{https://www.cosmos.esa.int/web/gaia/dpac/consortium}). Funding for the DPAC
has been provided by national institutions, in particular the institutions
participating in the {\it Gaia} Multilateral Agreement.
This publication makes use of data products from the Two Micron All Sky Survey, which is a joint project of the University of Massachusetts and the Infrared Processing and Analysis Center/California Institute of Technology, funded by the National Aeronautics and Space Administration and the National Science Foundation.
This research has made use of the NASA/IPAC Infrared Science Archive, which is funded by the National Aeronautics and Space Administration and operated by the California Institute of Technology.

\software{%
astropy \citep{price2018astropy}
matplotlib \citep{hunter07},
csv \citep{van1995python}, 
MOOG \citep{sneden73},
numpy \citep{vanderwalt11},
pandas \citep{mckinney2010data}, 
re \citep{chandra2015python}, 
requests \citep{chandra2015python}, 
scipy \citep{jones01}}

\bibliographystyle{aasjournal}
\bibliography{iuroederer,gc, sanilm}

\end{document}